\documentclass[aps,twocolumn,showpacs,superscriptaddress,nofootinbib]{revtex4}

\usepackage{graphicx}
\usepackage{latexsym}

\usepackage{color}

\begin{document}

\title{Rate in Template-directed Polymer Synthesis}

\author{Takuya Saito}
\email[Electric mail:]{saito@fukui.kyoto-u.ac.jp}
\affiliation{Fukui Institute for Fundamental Chemistry, Kyoto University, Kyoto 606-8103, Japan}

\def\Vec#1{\mbox{\boldmath $#1$}}
\def\degC{\kern-.2em\r{}\kern-.3em C}

\def\SimIneA{\hspace{0.3em}\raisebox{0.4ex}{$<$}\hspace{-0.75em}\raisebox{-.7ex}{$\sim$}\hspace{0.3em}} 

\def\SimIneB{\hspace{0.3em}\raisebox{0.4ex}{$>$}\hspace{-0.75em}\raisebox{-.7ex}{$\sim$}\hspace{0.3em}} 

\newcommand{\gsim}{\hspace{0.3em}\raisebox{0.5ex}{$>$}\hspace{-0.75em}\raisebox{-.7ex}{$\sim$}\hspace{0.3em}} 
\newcommand{\twoarrow}{\hspace{0.5em}\raisebox{0.25ex}{$\rightarrow$}\hspace{-1.5em}\raisebox{-.25ex}{$\leftarrow$}\hspace{0.5em} }

\date{\today}

\begin{abstract}
We discuss temporal efficiency of template-directed polymer synthesis, such as DNA replication and transcription, under a given template string.
To weigh the synthesis speed and accuracy on the same scale,
we propose a template-directed synthesis (TDS) rate, which contains an expression analogous to that for the Shannon entropy.
Increasing the synthesis speed accelerates the TDS rate, 
but the TDS rate is lowered if the produced sequences are diversified.
We apply the TDS rate to some production system models and investigate how the balance between the speed and the accuracy is affected by changes in the system conditions.
\end{abstract}

\pacs{05.40.-a,82.35.Pq,82.20.Hf}

\maketitle

\section{Introduction}
In biopolymer synthesis, sequence information is retrieved from a template string 
and stored in a polymer product.
Such template-directed polymer syntheses are exemplified by DNA replication and transcription, 
where catalysts such as polymerase enzymes develop the product polymer
along the template with the substrates serving as an energy source as well as material.
The fundamental mechanical characteristics have been extensively investigated from the standpoint of molecular motors~\cite{Science_Wang_1998,Nature_Wuite_2000}, and, moreover,
recent progress allows the monitoring of the real-time sequencing~\cite{Science_Eid_2009,Nature_Uemura_2010,NatureMethods_Flusberg_2010}.

One of the remarkable aspects in such polymerization is the selection of a substrate suitable for the template in the presence of thermal random forces, which may induce occasional mistakes.
Template-directed synthesis (TDS) including error occurrence has been explored to give a physical description~\cite{PNAS_Hopfield_1974,Biochimie_Ninio_1975,BioSystems_Bennett_1979,PNAS_Andrieux_2008,PNAS_Jarzynski_2008,PRL_Woo_2011,arXiv_Andrieux_2013}, 
with a particular focus on developing the nonequilibrium statistical thermodynamics for a sequence match~\cite{PNAS_Andrieux_2008,PNAS_Jarzynski_2008} and a scheme for lowering the error rate~\cite{PNAS_Hopfield_1974,Biochimie_Ninio_1975,BioSystems_Bennett_1979}.

In a biological system, a polymeric sequence is synthesized with a low error fraction, which is estimated as, e.g., $\sim 10^{-9}$ per nucleotides in {\it E. coli} genomic DNA~\cite{NatEdu_Pray_2008}.
A kinetic proofreading mechanism was proposed to achieve such remarkable fidelity~\cite{PNAS_Hopfield_1974,Biochimie_Ninio_1975}.
According to this scenario, a Michaelis scheme manages to reduce the error rate with an energy input relative to that in equilibrium.
Applying multiple steps of repeating these schemes is expected to increase the correctness of the terminal product. 
Here the following question arises:
Using multiple steps reduces the error rate, but does it greatly increase the necessary processing time?
That is, are the synthesis speed and the accuracy incompatible?
Even if an advantage in one conflicts with that of the other,
since the production speed and the accuracy are usually gauged by different scales, 
how should we recognize the balance between them?

Our purpose in the present article is to propose a TDS rate that weighs the speed and accuracy on the same scale to discuss the balance between the speed and the accuracy.
The TDS rate measures the performance in various kinds of the systems.
In the following sections, after giving the TDS rate definition, we build two models: (A) a jump process model and (B) a switching well potential model described by the Brownian dynamics.
On the basis of the TDS rate, we discuss the balance in these two cases.

\begin{figure}[b]
\begin{center}
\includegraphics{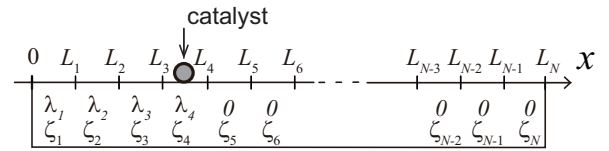}
\caption{
	Illustration of synthesizing a sequence string.
	Catalyst synthesizes a product polymer from the left end to the right end along a fixed template.
	  }
\label{fig1}
\end{center}
\end{figure}

\begin{figure}[b]
\begin{center}
\includegraphics{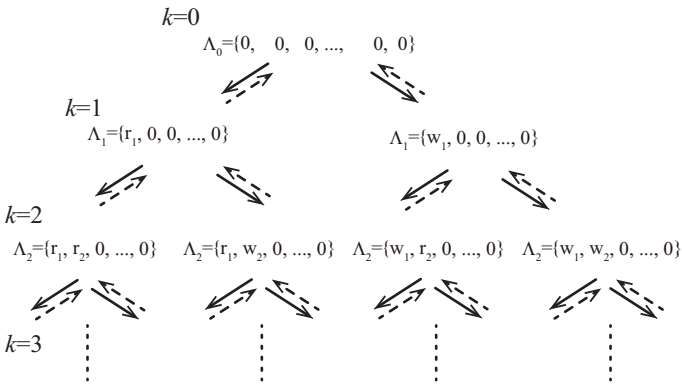}
\caption{
		Tree diagram of a binary system with binary element substrates $\lambda=r,w$ on mono element template $\zeta_1=\zeta_2=\cdots=\zeta_{N-1}=\zeta_N$.
		Right or wrong combinations correspond to $r$ or $w$, respectively.
	  }
\label{fig2}
\end{center}
\end{figure}

\section{Rate of template-directed synthesis}

Let us consider a TDS system.
A template string has $N$ code elements, which are arrayed at intervals of $l$ ($L_k\equiv kl$) along the $x$-axis as shown in Fig.~1.
The template sequence is $Z =\{ \zeta_1, \zeta_2, \cdots, \zeta_N \}$, where the elements denoted by $\zeta_k$ are ordered with respect to $k$ from $x=0$. 
Incorporated/incorporating substrate elements are denoted by $\lambda_k$ paired with template element $\zeta_k$.
The species number of the code elements is $M$, e.g., for DNA replication, $M=4$ and $\lambda, \zeta \in \{ {\rm A, T, G, C} \}$ (complementary base pairs are A-T and G-C).
In Fig.~1, a catalyst, such as DNA polymerase, is represented by the ball at $x$; the catalyst embeds the substrates $\lambda$ in the empty slots expressed by ``0", starting from the origin.
Incorporating $\lambda_k$, the entire substrate sequence is denoted by $\Lambda_{k} =\{ \lambda_1, \lambda_2, \cdots,\lambda_{k}, 0,\cdots,0 \}$.
The products are considered to be completed when the catalyst passes through $x=L_N$.
The terminal product has the full substrate sequence denoted by $\Lambda_{f} =\{ \lambda_1, \lambda_2, \cdots,\lambda_{N-1}, \lambda_{N} \}$.
After the completion, the catalyst is set to go back to $x=0$ with the empty sequence $\Lambda_{0} =\{ 0, 0, \cdots, 0,0\}$.
Through many trials, we obtain a set of the output probability flows $\{ J_x^{\Lambda_{f(1)}}, J_x^{\Lambda_{f(2)}}, J_x^{\Lambda_{f(3)}}, \cdots \}$.

Here, we discuss what products are more correct through examining the output probability flows.
To make the argument clear, let us take the simplest TDS system:
a binary system, where two substrates ($M=2$) are incorporated into a mono element template, $\zeta_1=\zeta_2=\zeta_3=\cdots=\zeta_{N-1}=\zeta_N$
(see Fig.~2).
One specific example is a system with substrates $\{$C,G$\}$ on template $Z=\{ \mathrm{G_1,G_2,G_3,}\cdots,\mathrm{G}_{N-1},\mathrm{G}_N \}$.
Conventionally, the combination C-G can be regarded as right, while G-G is regarded as wrong.
Hereafter, in accordance with right or wrong combinations, the substrates are expressed by $\lambda=r$ or $w$, respectively.
The catalyst repeats the same process $N$ times so that $2^N$ type sequences are eventually generated in the terminal products.
Regarding a physical meaning of right/wrong, we notice that majority or minor of the products should be considered as being right or wrong, respectively.
Thus, in this sense, the perfect right flow of the terminal products ($J_x^{\Lambda_{f}}$ of $\Lambda_{f} =\{ r_1, r_2, \cdots,r_{N-1}, r_{N} \}$) dominates the others.
Next, the products for which one wrong and $N-1$ right substrates are incorporated have the second largest flows.
Continuing, the flow deceases with increasing numbers of wrong elements $w_k$ embedded in the product sequence. 
Finally, $\Lambda_{f} =\{ w_1, w_2, \cdots,w_{N-1}, w_{N} \}$ is the least produced.

To advance the argument further, we can extend the ordering procedure for what products are more correct for general cases ($M\geq2$) beyond a binary choice.
The $M$-substrate system may finally produce $M^N$ type sequences.
If a synthesis system produces $M^N$ sequences at rates $\{ J_x^{\Lambda_{f(1)}} > J_x^{\Lambda_{f(2)}} > J_x^{\Lambda_{f(3)}}> \cdots > J_x^{\Lambda_{f(M^N)}} \}$, 
then the correct order is $\{ \Lambda_{f(1)}, \Lambda_{f(2)}, \Lambda_{f(3)}, \cdots, \Lambda_{f(M^N)} \}$, i.e., 
the sequence produced with a higher rate is recognized as being the more correct sequence.
Moreover, it is notable that, even if the template is not mono element, but arbitrary, a set of flows are observed, so that the same ordering procedure can be adopted as long as the sequence is fixed.

Next let us consider the system performance with a set of the flow $\{ J_x^{\Lambda_{f(1)}}, J_x^{\Lambda_{f(2)}},$ $\cdots, J_x^{\Lambda_{f(M^N)}} \}$ by developing the above arguments. 
The intuitive features of a TDS rate are expected to satisfy the following:

(i) 
A more biased flow distribution, e.g.,  $\{ J_x^{\Lambda_{f(1)}} \gg J_x^{\Lambda_{f(2)}} \gg$ $\cdots \gg J_x^{\Lambda_{f(M^N)}} \}$ is evaluated as a better TDS rate. 
In the limiting conditions, the maximum TDS rate is achieved if a single sequence is generated. 
In contrast, if all sequences are synthesized with the same product rate, the TDS rate is minimized.

(ii) If the ratio of synthesized sequences in the distribution is maintained constant, the TDS rate is proportional to the total product speed, i.e., it is an intensive quantity.

To meet conditions (i) and (ii), the TDS rate ${\cal I}_J$ is defined in terms of $J_x^{\Lambda_{f}}$ as
\begin{eqnarray}
{\cal I}_J 
&\equiv& 
k_\mathrm{B}T \sum_{\Lambda_{f}}  J_x^{\Lambda_{f}} \log{\frac{J_x^{\Lambda_{f}}}{ (J_x/M^N) }}.
\label{infoflow_overd}
\end{eqnarray}
where $k_\mathrm{B}$ is the Boltzmann constant and $T$ is the absolute temperature (see appendix A for the other difinition).
At the fixed total probability flow $J_x\equiv \sum_{\Lambda_{f}}  J_x^{\Lambda_{f}}$, ${\cal I}_J$ defined by eq.~(\ref{infoflow_overd}) satisfies $0 \leq {\cal I}_J \leq k_\mathrm{B}T J_x \log{M^N}$.
Condition (ii) clearly holds in eq.~(\ref{infoflow_overd}).
To confirm that condition (i) holds,
let us first consider the best/worst ways to allocate $J_x^{\Lambda_{f}}$ while maintaining the total flow $J_x$ unchanged.
Equation~(\ref{infoflow_overd}) contains a functional form analogous to that of the Shannon entropy~\cite{Shannon_1949,Cover_Thomas} with respect to the flow $J_x^{\Lambda_{f}}$ instead of the probability.
The rate is minimized to ${\cal I}_J=0$ by taking an even distribution over all the possible sequences: $J_x^{\Lambda_{f(1)}} = J_x^{\Lambda_{f(2)}}=J_x^{\Lambda_{f(3)}}=\cdots=J_x/M^N$.
In contrast, the function is maximized to ${\cal I}_J = k_\mathrm{B}T J_x \log{M^N}$ if only one sequence is generated, e.g., $J_x^{\Lambda_{f(1)}}=J_x$ with $J_x^{\Lambda_{f(2)}}=J_x^{\Lambda_{f(3)}}=\cdots=0$.
Next, to consider intermediate conditions, let us compare two different TDS systems, 
each of which has three sequence products: $\{ \Lambda_{f(1)}, \Lambda_{f(2)}, \Lambda_{f(3)} \}$ or $\{ \Theta_{f(1)}, \Theta_{f(2)}, \Theta_{f(3)} \}$. 
Those flows satisfy the magnitude relation
\begin{eqnarray}
\begin{array}{ccccc}
\{ J_x^{\Lambda_{f(1)}} & > & J_x^{\Lambda_{f(2)}} & > & J_x^{\Lambda_{f(3)}} \}  \\
\rotatebox{90}{$=$}      &   & \rotatebox{90}{$<$}    &   & \rotatebox{90}{$>$}   \\
\{ J_x^{\Theta_{f(1)}}  & > & J_x^{\Theta_{f(2)}}  & > &  J_x^{\Theta_{f(3)}} \}, \\
\end{array} 
\end{eqnarray} 
where each system has the same total flows $J_x=\sum_{\Lambda_{f}}J_x^{\Lambda_{f}}=\sum_{\Theta_{f}}J_x^{\Theta_{f}}$.
The largest flows in the respective systems are the same, in that $J_x^{\Lambda_{f(1)}}=J_x^{\Theta_{f(1)}}$, but the second largest one in the former system is larger than its counterpart, as $J_x^{\Lambda_{f(2)}}>J_x^{\Theta_{f(2)}}$.
In the criterion of eq.~(\ref{infoflow_overd}), ${\cal I}_J$ gives the higher rate to $\{ \Lambda_{f(1)}, \Lambda_{f(2)}, \Lambda_{f(3)} \}$ with the second largest flow $J_x^{\Lambda_{f(2)}}$ rather than to the other system.
Indeed, the terminal products in $\{ \Lambda_{f(1)}, \Lambda_{f(2)}, \Lambda_{f(3)} \}$ are more concentrated on the two sequences $\Lambda_{f(1)}$, $\Lambda_{f(2)}$, 
indicating that the products have a narrower distribution over sequences.
Thus, by using the sequence distribution, the TDS rate can organize the order in general production systems, which meets condition (i).

\section{Model A: Jump process model}

We will next apply the TDS rate to specific models.
Let us first give a simple jump process model in the binary system $r$ or $w$ like in Fig.~2.
For simplicity, we consider that the catalyst at the $k$-th site only jumps forward by $l$ to the next, i.e., $(k+1)$-th, site, while ignoring the backward process.
The forward step accompanies the incorporating process. 
The jump transition rate of right or wrong substrates is $W^{r+}$ or $W^{w+}$, respectively, which are both independent of the past incorporated substrates.
The probability at the $k$-th site is denoted by $P^{\Lambda_k}=P^{\{\lambda_1,\cdots, \lambda_k, 0, \cdots,0 \}}$ with $\lambda_n=r_n,w_n$.
At each site, the time evolution is governed by
\begin{eqnarray}
\partial_t P^{\{\lambda_1,\cdots, \lambda_{k-1}, r_k,0, \cdots,0 \}}
&=& W^{r+} P^{\{\lambda_1,\cdots, \lambda_{k-1},0, \cdots,0 \}}
\nonumber \\
&& -W P^{\{\lambda_1,\cdots, \lambda_{k-1},r_k, \cdots,0 \}}
\nonumber \\
\partial_t P^{\{\lambda_1,\cdots, \lambda_{k-1}, w_k,0, \cdots,0 \}}
&=& W^{w+} P^{\{\lambda_1,\cdots, \lambda_{k-1},0, \cdots,0 \}}
\nonumber \\
&& -W P^{\{\lambda_1,\cdots, \lambda_{k-1},w_k, \cdots,0 \}}
\nonumber \\
\label{t_model_A}
\end{eqnarray}
where $W\equiv W^{r+}+W^{w+}$. 
At the end of the synthesis, the transition $\Lambda_f \rightarrow \Lambda_0$ with the rate $W$ takes place.
When looking at the substrate elements at the $k$-th site, the total probabilities that $\lambda_k=r_k$ or $w_k$ are incorporated are given by, respectively,
\begin{eqnarray}
P^r &\equiv& \sum_{\{\lambda_1, \cdots, \lambda_{k-1} \}} P^{\{\lambda_1,\cdots, \lambda_{k-1}, r_k,0, \cdots,0 \}},
\nonumber \\
P^w &\equiv& \sum_{\{\lambda_1, \cdots, \lambda_{k-1} \}} P^{\{\lambda_1,\cdots, \lambda_{k-1}, w_k,0, \cdots,0 \}}.
\label{hmg_model_A}
\end{eqnarray}
We then assume the steady state ($\partial_t P^r=\partial_t P^w=0$) and homogeneous substrate distribution
so that $P^r+P^w = \sum_{\{\lambda_1, \cdots, \lambda_{k-1} \}} P^{\{\lambda_1,\cdots, \lambda_{k-1},0, \cdots,0 \}}$.
Taking summation over $\{ \lambda_1, \cdots, \lambda_{k-1} \}$ in eq.~(\ref{t_model_A}),
we have
\begin{eqnarray}
\partial_t P^r = W^{r+}(P^r+P^w)-WP^{r}
\nonumber \\
\partial_t P^w = W^{w+}(P^r+P^w)-WP^{w}.
\label{Pr_Pw}
\end{eqnarray}
Setting $\partial_t P^r=\partial_t P^w=0$ into eq.~(\ref{Pr_Pw}),
we arrive at a stationary solution $P^r = {\cal A} W^{r+}$ and $P^w = {\cal A} W^{w+}$ with ${\cal A}$ being the normalization factor.
Using $P^r$ and $P^w$, we finally find that each terminal product is generated with the probability
\begin{eqnarray}
P^{\Lambda_f} = \frac{(W^{r+})^{n_r}(W^{w+})^{N-n_r}}{N W^N},
\label{product_model_A}
\end{eqnarray}
where $n_r$ is the number of right substrates incorporated into the product and the probability is normalized in the summation over the terminal sequences $\sum_{\Lambda_f}P^{\Lambda_f}=1/N$.
In this case, each flow is given by $J_x^{\Lambda_f}=WP^{\Lambda_f}$, and the criterion that the majority flow is correct defines $W^{r+}>W^{w+}$.
Indeed, from the above, we can confirm that a decrease in $n_r$ reduces $J_x^{\Lambda_f}$, organizing the flow ordering.

Next applying eq.~(\ref{product_model_A}) to the TDS rate, we have
\begin{eqnarray}
{\cal I}_J
&=&
\sum_{\Lambda_f} J_x^{\Lambda_f} \log{\left(\frac{J_x^{\Lambda_f}}{J_x/2^N}\right)} 
\label{TDS_model_A}\\ 
&=& \sum_{n_r=0}^N {}_NC_{n_r} Wl \frac{(W^{r+})^{n_r}(W^{w+})^{N-n_r}}{NW^N} 
\nonumber \\ 
&& \quad \times \log{\left[ \frac{(W^{r+})^{n_r}(W^{w+})^{N-n_r}}{W^N/2^N} \right]}
\nonumber \\
&=& lW^{r+} \log{\left( \frac{W^{r+}}{W/2} \right)} +lW^{w+}\log{\left( \frac{W^{w+}}{W/2} \right)}.
\label{result_model_A}
\end{eqnarray}
Equation~(\ref{result_model_A}) is written as a function of only the effective velocities $W^{r+}l$ and $W^{w+}l$. 
This means, if the backward process is negligible, then the TDS rate is estimated by observing these two velocities.

In addition, conditions (i) and (ii) introduced as the TDS rate definition can be confirmed in eq.~(\ref{TDS_model_A}):
For a fixed $W$, the TDS rate is maximized when ${\cal I}_J=NlW \log{2}$ with $W=W^{r+}$.
In contrast, it is minimized when ${\cal I}_J=0$ with $W^{r+}=W^{w+}=W/2$.
Moreover, an additional meaning of the TDS rate is found, as follows.
Let us increase $W^{w+}$ while keeping $W^{r+}$ constant.
The flow $J_x=Wl/N$ is clearly enhanced, but the TDS rate decreases because $\partial {\cal I}_J/\partial W^{w+}<0$ under $W^{r+}>W^{w+}$.
This means that, even if the correct product is produced at the same rate, the contamination of the wrong products eventually diminishes the TDS rate.

\begin{figure}[t]
\begin{center}
\includegraphics[scale=0.75]{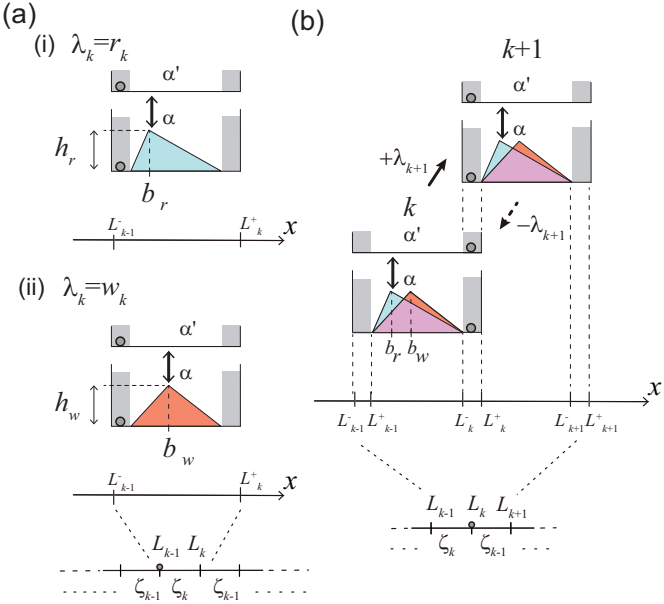}
      \caption{(Color online) 
	  Schematic representation of switching well potentials in a binary system model. 
	  Each of the elements ($\lambda_k =r_k,w_k$) has the two internal states $\alpha, \alpha'$.
	  (a) Right or wrong combinations have (i) blue or (ii) red shaded potentials, respectively.
	  In the internal state $\alpha$, the peaks satisfy $b_r<b_w$, $b_r<l/2$, and $h_r=h_w$.
	  (b) Sequence state transition between the $k$-th and $(k+1)$-th elements. 
	  In the image, right/wrong potentials are shown.
	  }
\label{fig3}
\end{center}
\end{figure}

\begin{figure}[t]
\begin{center}
\includegraphics{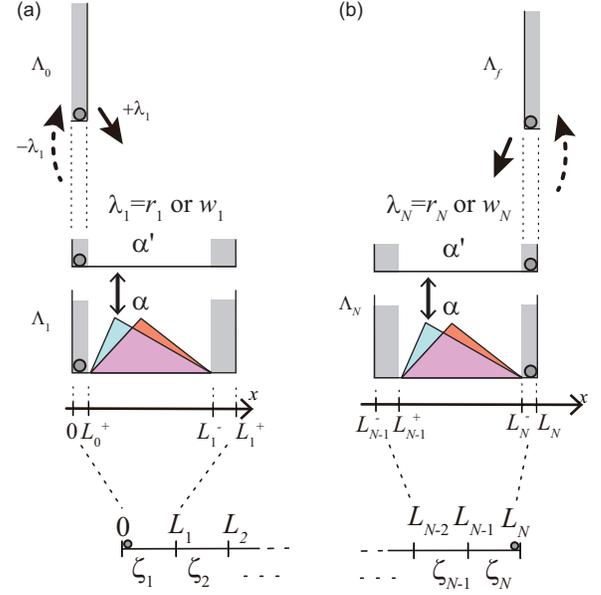}
      \caption{(Color online) 
	  Schematic representation of switching well potentials around the boundaries.
	  Sequence state transition (a) between all empty and incorporating a first element or 
	  (b) between synthesizing last element and full.
	  In the image, right/wrong potentials are shown.
	  }
\label{fig4}
\end{center}
\end{figure}

\section{Model B: Sequential well potential model}

The catalysts in this model, such as polymerase, are molecular motors consuming the energy source through the TDS reaction. 
In this section, we examine how the TDS rate is associated with the energy input by describing the system in terms of Brownian motion.
Again, we adopt a binary system like that in Fig.~2.

Let us look at the step at which the catalyst synthesizes the $k$-th substrate.
Each incorporating process is captured by the Brownian motion in a well potential as in Fig.~3\,(a).
The potential is indicated by the gray regions on the two sides and the catalyst is first in the left gray region.
Completing each step is described by a transfer process from the left- to the right-hand gray regions (see appendix B for general cases). 

The potential shape in the middle region reflects the affinity of the $\lambda_k$-$\zeta_k$ combination relying on what substrates (i) $\lambda_k=r_k$, (ii) $\lambda_k=w_k$ are caught in the $k$-th region.
A motion driven by the consumption of the energy supply is introduced here by applying a fluctuating ratchet model~\cite{Adjari_Prost_1992,PRL_Magnasco_1993,PRL_Astumian_1994,PRL_Doering_1994,EPL_Bartussek_1994,PRL_Prost_1994,EPL_Parrondo_1998,PhysRep_Reimann_2002,arXiv_Horowitz_2013}.
In particular, utilizing the idea of a flashing ratchet model~\cite{Adjari_Prost_1992,PRL_Magnasco_1993,PRL_Astumian_1994,PRL_Doering_1994,EPL_Bartussek_1994,PRL_Prost_1994,EPL_Parrondo_1998,PhysRep_Reimann_2002}, 
the potentials are switched accompanied by the transition of the internal state between the flat ($\alpha'$: top figure) and triangle ($\alpha$: bottom figure) both for (i) $r_k$ and for (ii) $w_k$ in Fig.~3\,(a). 
Note that the transition takes place stochastically with the rate $W^{\Lambda_{k,\alpha} \twoarrow \Lambda_{k,\alpha'}}(x)$, and the same substrate $\lambda_k$ is maintained before and after the transition $\alpha \twoarrow \alpha'$.
The potential shapes are exactly described by
\begin{eqnarray}
&&U^{\Lambda_{k,\beta}}(x) 
=
\left\{ \begin{array}{ll}
+\infty  & (x \leq L_{k-1}^- ) \\
0 & (L_{k-1}^-<x \leq L_{k-1}^+ ) \\
\mathrm{see~eq.~(\ref{flash})}  & (L_{k-1}^+\leq x \leq L_k^-) \\
0 & (L_k^-\leq x<L_k^+) \\
+\infty  & (L_k^+ \leq x) 
\end{array} \right. 
\nonumber \\
\end{eqnarray}
where $\Lambda_{k,\beta} =\{ \lambda_1, \lambda_2, \cdots,\lambda_{k-1}, \lambda_{k}, 0,0, \cdots 0,0 \}_\beta$ is the substrate sequence with internal state $\beta=\alpha,\alpha'$, $L_k^-\equiv L_k-\Delta l/2$, and $L_k^+\equiv L_k+\Delta l/2$.
The potentials in the switching region are
\begin{eqnarray}
U^{\Lambda_{k,\alpha}}(x) 
&=& 
\left\{ 
\begin{array}{ll}
& a_{1,\lambda}(x-L_{k-1}^+)    \\
& \qquad \cdots (L_{k-1}^+<x<L_{k-1}^+ +b_{\lambda}) \\
\nonumber \\
&-a_{2,\lambda}(x-(L_{k-1}^+ +b_{\lambda})) +a_{1,\lambda}b_{\lambda} \\
& \qquad \cdots (L_{k-1}^+ +b_{\lambda} <x<L_k^-)
\end{array} \right. 
\nonumber \\
\uparrow \downarrow &&
\nonumber \\
U^{\Lambda_{k,\alpha'}}(x) 
&=& 0, 
\label{flash}
\end{eqnarray}
where $a_{1,\lambda}$ and $a_{2,\lambda}$ are numerical constants, $b_\lambda$ determines the peak's position in the triangle potential, and $a_{2,\alpha}=a_{1,\lambda}b_\lambda/(L_k^--L_{k-1}^+-b_\lambda)$ is adopted to connect the profile continuously.
In the asymmetric shape $b_\lambda<l/2$ with $a_{1,\lambda}>0$ and $a_{2,\lambda}>0$, 
the catalyst may be driven from left to right gray regions~\cite{Adjari_Prost_1992,PRL_Magnasco_1993,PRL_Astumian_1994,PRL_Doering_1994,EPL_Bartussek_1994,PRL_Prost_1994,EPL_Parrondo_1998,PhysRep_Reimann_2002}.
The difference between $\lambda=r$ and $\lambda=w$ is the peak's position, and $b_r<l/2$ and $b_r<b_w$ are adopted with the same other parameters ($h_r=h_w$, etc.), so that a rightward motion $\lambda=r$ is larger than that of $\lambda=w$.
Thus, the notation regarding $\lambda=r,w$ is reasonable
from the criterion that the majority is correct.

Further, the potential switching to the neighboring sequences (Fig.~3\,(b)) builds up a series of synthesizing steps.
Being in the $k$-th right-hand gray region of $U^{\Lambda_{k,\beta}}(x)$,
the catalyst can catch the new substrate and transit to the $(k+1)$-th left-hand gray region of $U^{\Lambda_{k+1,\delta}}(x)$,
where
\begin{eqnarray}
\Lambda_{k,\beta} &=& \{ \lambda_1, \cdots, \lambda_{k-1}, \lambda_k, ~~0, ~~\,0, \cdots, 0 \}_\beta 
\nonumber \\
\Lambda_{k+1,\delta} &=&\{ \lambda_1, \cdots, \lambda_{k-1}, \lambda_k, \lambda_{k+1}, 0, \cdots, 0 \}_{\delta}.
\nonumber 
\end{eqnarray}
and the new internal states ($\delta=\alpha,\alpha'$) at the $(k+1)$-th site are stochastic.
On the other hand, the detaching substrate process is represented by the converse transition.
The event occurs with the transition rate $W^{\Lambda_{k,\beta} \twoarrow  \Lambda_{k+1,\delta}}(x)$ corresponding to the catching/detaching rates.

To begin and close the entire synthesizing process, the starting/terminal potentials around boundaries are added as shown in Figs.~4\,(a)/(b).
\begin{eqnarray}
U^{\Lambda_0}(x) 
= 
\left\{ \begin{array}{ll}
0  & (0 \leq x < L_0^+) \\
+\infty  & (L_0^+ \leq x) \\
\end{array}, \right.
\end{eqnarray}
\begin{eqnarray}
&&U^{\Lambda_{f}}(x) 
=
\left\{ \begin{array}{ll}
+\infty  & (x \leq L_N^- ) \\
0  & (L_N^-<x \leq L_N) \\
\end{array} \right.
\end{eqnarray}
Similarly, the potentials are switched with the rates 
$W^{\Lambda_0 \twoarrow \Lambda_{1,\alpha}}(x)$ and $W^{\Lambda_{N,\alpha} \twoarrow \Lambda_{f}}(x)$.

\begin{figure}[t]
\begin{center}
\includegraphics{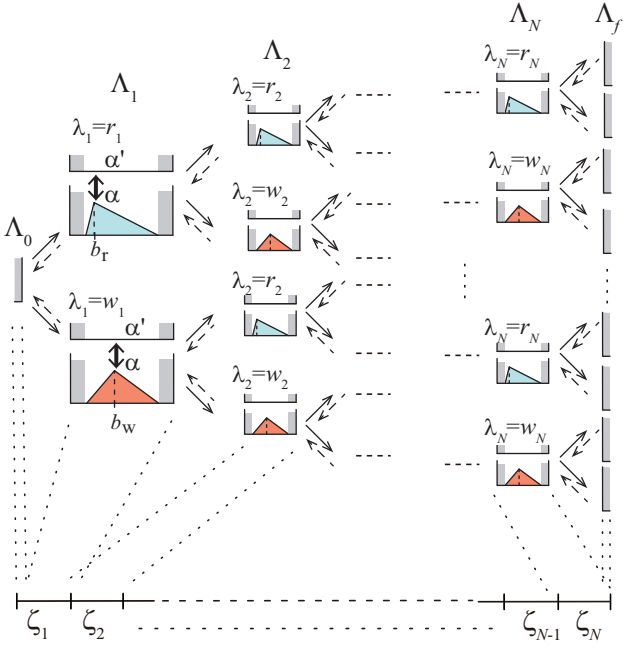}
      \caption{(Color online) 
	  Entire possible process of switching well potentials in a binary system model. 
	  The paths traced by arrows correspond to the tree diagram in Fig.~2.
	  }
\label{fig5}
\end{center}
\end{figure}

\begin{figure}[t]
\begin{center}
\includegraphics{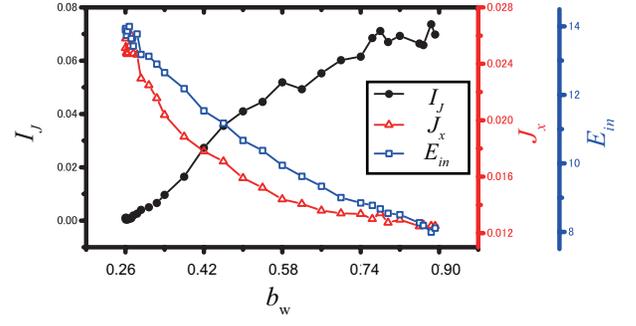}
      \caption{(Color online) 
	  Numerical result of the TDS rate $\cal{I}_J$, total probability flow $J_x$, and energy input rate for the binary system model.
	  The calculation is performed using the underdamped Langevin dynamics.
	   The abscissa axis is $b_\mathrm{w}$ and $b_\mathrm{r}$ is held constant at 0.26.
	  The other parameters are set as $k_\mathrm{B}T=4.2$, $N=3$, $l=1.0$, $\Delta l=0.10$, $\gamma=100$, $m=0.1$, $\Delta U=0$, $\Delta t=10^{-8}$, $W_{\Lambda_{k,\beta} \rightarrow \Lambda_{k+1,\delta}}=W_{\Lambda_0 \rightarrow \Lambda_{1,\beta}}=50$, $W_{\Lambda_{k+1,\beta} \rightarrow \Lambda_{k,\delta}}=W_{\Lambda_{1,\beta} \rightarrow \Lambda_0}=5$, $W_{\Lambda_{k,\alpha} \rightarrow \Lambda_{k,\alpha'}}=W_{\Lambda_{k,\alpha'} \rightarrow \Lambda_{k,\alpha}}=20 \times (1-W_{\Lambda_{k+1,\beta} \rightarrow \Lambda_{k,\delta}}\Delta t)$, $W_{\Lambda_{N,\alpha} \rightarrow \Lambda_f}=200$, $W_{\Lambda_f \rightarrow \Lambda_{N,\alpha}}=50$, and $h_r=h_w=50.0$.
	  }
\label{fig6}
\end{center}
\end{figure}

Having defined all steps, we here overview the entire process ($\Lambda_0 \twoarrow \Lambda_1 \twoarrow \Lambda_2 \twoarrow \cdots \twoarrow \Lambda_{N-2} \twoarrow \Lambda_{N-1} \twoarrow \Lambda_{f}$) shown in Fig.~5, which corresponds to the tree diagram in Fig.~2.
As in Fig.~4\,(a), the catalyst enters at $x=0$ in $\Lambda_0=\{0,0, \cdots, 0\}$, and stochastically transfers to a next neighbor potential $\Lambda_1=\{r_1,0,0, \cdots, 0\}$ or $\{w_1,0,0, \cdots, 0\}$ according to incoming substrates.
Next, as in Fig.~3\,(a), it may be driven by switching well potentials.
If arriving at the right-hand gray region in $\Lambda_1=\{\lambda_1,0, \cdots, 0\}$ with $\lambda_1=r_1,\,w_1$, it stochastically transfers to a next $\Lambda_2=\{\lambda_1,r_2,0, \cdots, 0\}$ or $\{\lambda_1,w_2,0, \cdots, 0\}$ as in Fig.~3\,(b).
After that, if the qualitatively same processes (Figs.~3\,(a),(b)) repeat, a forward process develops $\rightarrow \Lambda_3 \rightarrow \Lambda_4 \rightarrow \cdots \rightarrow \Lambda_{N}$. 
Finally, the catalyst arrives at the right-hand gray region of the last element $\Lambda_{N}$, where the potential is stochastically switched into that of $\Lambda_{f}$ as in Fig.~4\,(b).
Passage through $x=L_N$ completes the entire synthesizing process.

In the analyses, the steady state is assumed:
the product sequence $\Lambda_{f}$ is still retained just before $x=L_N$.
Passing across $x=L_N$, the catalyst goes back to the starting point $x=0$, at which the sequences are initialized, i.e., $\Lambda_{f} \Rightarrow \Lambda_0$. 
On the other hand, if it moves back from $x=0$ to $x=L_N$, the last produced sequence is recovered, $\Lambda_0 \Rightarrow \Lambda_{f}$.

We demonstrate a numerical results of the switching potential model introduced in Figs.~3--5. 
The TDS rate ${\cal I}_J$, the total flow $J_x$, and the energy input rate $E_{\rm in}$ are plotted as functions of the wrong element's peak position $b_w$, with the right element's one held fixed at $b_r=0.26$ (see Fig.~6). 
If the peak position $b_w$ is close to $b_r$, then the total probability flow is high but the TDS rate is low.
In the limit $b_w \rightarrow b_r$, the TDS rate should exhibit ${\cal I}_J \rightarrow 0$, because ``$r_k$"/``$w_k$" become indistinguishable.
In contrast, as $b_w$ becomes more distant from that of $b_r$, the probability flow decreases but the TDS rate increases overall.
Note that this model limits the TDS rate to not more than ${\cal I}_J^0\equiv k_BTJ_x\log{2^N}$ due to the incomplete inhibition of the wrong element's flow.
The above trends read from the plots in Fig.~6 indicate that a high production flow does not mean a high TDS rate.

In addition, in Fig.~6, the energy input rate increases when letting $b_w \rightarrow b_r$, although the TDS rate shrinks.
This indicates that a high energy input rate does not necessarily correspond to a high TDS rate.
However, if the energy input is properly consumed, the TDS rate is expected to increase as the energy input rate is increased.
To find what factors disturb the TDS rate, 
we decompose the TDS rate through the Fokker--Planck equation with the discrete state transition.
The probability density function is governed~\cite{vanKampenBook,GardinerBook,RiskenBook,SekimotoBook} by
\begin{eqnarray}
 \partial_t P^{\Lambda}(x)
+\partial_x J_x^{\Lambda}(x)
&=&\sum_{\Lambda'} [W_{\Lambda' \rightarrow \Lambda}(x) P^{\Lambda'} 
-W_{\Lambda \rightarrow \Lambda'}(x) P^{\Lambda}],
\nonumber
\end{eqnarray}
\begin{eqnarray}
J_x^{\Lambda}(x)
= \frac{1}{\gamma} \left[ -\frac{\partial U^{\Lambda}(x)}{\partial x}P^{\Lambda}(x) -k_\mathrm{B}T \frac{\partial P^{\Lambda}(x)}{\partial x} \right].
\label{over_FP}
\end{eqnarray}
where the $\Lambda_{0},\Lambda_{k,\alpha},$ and $\Lambda_{f} \rightarrow \Lambda$ notation is omitted, $\gamma$ is the viscous friction coefficient, and the probability of being in the $\Lambda$ state is denoted by $P^{\Lambda}$.
Using the steady state condition $\partial_tP(x)^\Lambda=0$ and the continuity of the total probability density $\sum_{\Lambda,\Lambda'} [W_{\Lambda' \rightarrow \Lambda}(x) P^{\Lambda'} 
-W_{\Lambda \rightarrow \Lambda'}(x) P^{\Lambda}]=0$, the TDS rate is decomposed to
\begin{eqnarray}
{\cal I}_J
&=&
E_{\rm in}
-T \left< \dot{s}_{\rm seq} \right>
-\Psi
+{\cal I}_J^0,
\label{main_overdamped}
\end{eqnarray}
where 
\begin{eqnarray}
E_{\rm in}&=&\int_0^{L_N} \mathrm{d}x \sum_{\Lambda,\Lambda'} [U^{\Lambda} -U^{\Lambda'}] W_{\Lambda' \rightarrow \Lambda} P^{\Lambda'}, 
\nonumber \\
\left< \dot{s}_{\rm seq}\right>&=&k_\mathrm{B} \int_0^{L_N} \mathrm{d} x~ \sum_{\Lambda,\Lambda'} W_{\Lambda \rightarrow \Lambda'}P^{\Lambda} \log{(P^{\Lambda}/P^{\Lambda'})}, 
\nonumber \\
\Psi&=&\sum_{\Lambda}\int_0^{L_N} \mathrm{d}x~  \gamma P^{\Lambda} \left( \frac{ J_x^{\Lambda} }{P^{\Lambda}} \right)^2,
\nonumber \\
{\cal I}_J^0 &\equiv& k_\mathrm{B}T J_x\log{ 2^N }.
\label{decomp_over}
\end{eqnarray}
In eq.~(\ref{decomp_over}),
$E_{\rm in}$ is the input energy rate, $\left< \dot{s}_{\rm seq} \right>$ is called the stochastic entropy change rate in the sequence transition, $\Psi$ is a non-negative term, and ${\cal I}_J^0$ is the base flow, which is a linear flow term.
The above decomposition was 
obtained in the form of the time derivative calculation of the H-theorem with some additional potential conditions close to the boundary
(see the appendix C).
Here, let us look at the right-hand side of eq.~(\ref{main_overdamped}).
While $\left< \dot{s}_{\rm seq} \right> $ is zero in the equilibrium condition, it should be positive in a typical TDS system.
The last two $T\left< \dot{s}_{\rm seq} \right>+\Psi$ lead to a negative contribution to ${\cal I}_J$ in eq.~(\ref{main_overdamped}), while the increase of the energy input rate $E_\mathrm{in}$ boosts the TDS rate.
In total, the TDS rate ${\cal I}_J$ is essentially determined by the distance between the energy input $E_{\rm in}$ and the negative contributions $T\left< \dot{s}_{\rm seq} \right>+\Psi$.
Therefore, to improve the TDS rate, a high energy input is not necessarily the proper way, but rather a better balance may need to be found.

\section{Discussion}

So far, we have considered polymer synthesis.
Here, let us consider the application of the TDS rate to other systems.
DNA sequencing is a hard task in terms of labor hours, as well as cost.
Recently, investigations have been carried out to develop a rapid sequencing device that utilizes driven polymer translocation, which is the polymer passage through a pore induced by a voltage drop across the pore~\cite{JPolym_Fyta_2011}.
Moreover, the sequences are read by applying a voltage across the pore on a plate and by monitoring the temporal change of electric current. 
The experimental results provide the probability $P_{\Lambda_{f}} (\tau)$, where $\tau$ is the passage time and $\Lambda_{f}$ is the $N$ sequences read with the current. 
In this case, the rate in the sequencing is quantified as
\begin{eqnarray}
\frac{{\cal I}_J}{k_\mathrm{B}T}
=
\int_0^{+\infty} d\tau \sum_{\Lambda_{f}} \frac{P_{\Lambda_{f}} (\tau)}{\tau}
 \log{ \frac{P_{\Lambda_{f}}(\tau)/\tau}{(P(\tau)/\tau)/ M^N }},
\label{translocation}
\end{eqnarray} 
where the polymeric probability flow corresponds to $P_{\Lambda_{f}}(\tau)/\tau$.
All quantities are experimentally measurable and one idea is to evaluate the device performance by using the temporal rate (eq.~(\ref{translocation})).

Polymeric chain reaction (PCR) has been used as a basic molecular biology method.
Denaturation, annealing, and elongation processes repeat to achieve the exponential growth of the product in the PCR cycle.
A complementary DNA string is created in the elongation.
Controlling the physicochemical condition, e.g., changing dNTP concentration, should affect the required elongation time.
Because the system develops as an exponential growth system, 
we here propose ${\cal I}_J(\tau)/k_\mathrm{B}T= \sum_{\Lambda_{f}} (P_{\Lambda_{f}}/\tau) \log{[ (P_{\Lambda_{f}}/\tau) / [(P/\tau) M^N)] ]}$, with cycle time $\tau$ and probability of a synthesized $N$-sequence product $P_{\Lambda_{f}}$.
The cycle is externally controlled in the experiment so that
comparing the rates at different times $\tau$ or under various conditions makes it possible to find better experimental parameters.

\section{Summary}
In this article, we propose the TDS rate, which measures the synthesis speed and the sequence convergence on the same scale. 
The TDS rate is found to contain an expression analogous to that for the Shannon entropy by replacing the probability densities with the probability flows. 
The proposed TDS rate provides the ordering direction of the system performances, which takes account of the speed and accuracy.
If diversified sequences are produced, then the TDS rate is low, while it is increased by accelerating the production speed. 
In particular, in extreme examples, the TDS rate is maximized as the product sequences converge on a single sequence.
In contrast, the rate is minimized by setting the product flows to be evenly distributed over all possible sequences.
The rate definition is expected to be used to quantify the temporal accuracy in a TDS system or, for example, a DNA sequencing device utilizing polymer translocation.

We then construct TDS system models and examine how the TDS rate changes with the system conditions.
The results suggest that a high production flow does not mean a high TDS rate.

\acknowledgments
The author thanks T.\ Sakaue of Kyushu University and R.\ Okamoto of Kyoto University for useful discussions.

\subsection*{APPENDIX A: Underdamped case}
The underdamped case can be given a different TDS rate definition from eq.~(\ref{infoflow_overd}).
This case can involve a momentum integral as ${\cal I}_J \equiv k_\mathrm{B}T \int_{-\infty}^{+\infty} \mathrm{d}p
 \sum_{\Lambda_f}  J_x^{\Lambda_f} \log{\left[ J_x^{\Lambda_f} / (J_x/M^N)  \right]}$ with the flows $J_x^{\Lambda_f}(p)$ passing through $x=L_N$.

 \begin{figure}[b]
\begin{center}
\includegraphics{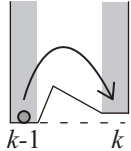}
      \caption{
	  Schematic representation of the case in which the internal energy gets higher through the synthesis step.
	  }
\label{fig7}
\end{center}
\end{figure}
 
\section*{APPENDIX B: General switching well potential}

In this appendix, we generalize the well potential model to cases for $M$ substrates on an arbitrary template with $M'$ elements.
It is notable that, even in this situation, the total sequence number of the terminal products is $M^N$ by counting the order of the substrate sequences at the fixed template sequence.

If the well potentials do not depend on the incorporated element history, the potential shape is determined only by the present incorporating element.
In the $k$-th element region of $\Lambda_k=\{ \lambda_1,\lambda_2,\cdots,\lambda_{k-1},\lambda_{k},0,\cdots,0 \}$,
the $M$ potentials $U^{\Lambda_{k(1)}},U^{\Lambda_{k(2)}},\cdots,U^{\Lambda_{k(M)}} $ may appear by coupling the template element with the $M$ substrate elements.
Then, in total, the $M'$ template elements with $M$ substrates create the $M' \times M$ well potentials in the TDS system. 
Note that, if adding the internal states, more potentials appear.

In addition, we can add the internal energy stored in the products, which means $U^{\Lambda_k}(L_{k-1})\neq U^{\Lambda_k}(L_{k})$. 
The height of the right-hand gray region is usually higher than that of the left-hand one $U^{\Lambda_k}(L_{k-1})<U^{\Lambda_k}(L_{k})$ because of the binding energy in the products (see Fig.~7). 
In the practical case, $U^{\Lambda_k(1)}(L_{k})=U^{\Lambda_k(2)}(L_{k})=U^{\Lambda_k(3)}(L_{k})=\cdots$ is expected since the terminal products are typically similar.
Such model extension does not alter the qualitative argument in the main text.
In some situations, however, we might have to take account of the difference $\Delta U^{\Lambda_{f}} \neq \Delta U^{\Lambda_{f}'}$ between the distinct sequences ($\Lambda_{f} \neq \Lambda_{f}'$).
If we are taking account of this difference, then we should include the product stability associated with the polymeric string breakage.

\subsection*{APPENDIX C: Relevance to energy input}
This section describes how the TDS rate is associated with the energy input.
Before the main issue, let us confirm the energy conservation as discussed in \cite{JPhysSocJPN_Yoshimori_2012}.
From $\sum_{\Lambda} \int \mathrm{d}x U^\Lambda(x) \partial_t P^{\Lambda}=0$, we have
\begin{eqnarray}
E_{\rm in}&=&\int_0^{L_N} \mathrm{d}x \sum_{\Lambda,\Lambda'} [U^{\Lambda} -U^{\Lambda'}] W_{\Lambda' \rightarrow \Lambda} P^{\Lambda'}
\nonumber \\
E_{\rm str}&=&\sum_{\Lambda_{f}} (U^{\Lambda_{f}}(L_N)-U^{\Lambda_0}(0))J_x^{\Lambda_{f}}(L_N)
\nonumber \\ 
\frac{\left<\mathrm{d}' Q\right>}{\mathrm{d}t} &=& -\sum_{\Lambda} \int_0^{L_N} \mathrm{d}x ~f^{\Lambda} J_x^{\Lambda}, 
\label{energy_in_over} 
\end{eqnarray}
where $E_{\rm str}$ is the internal energy rate stored in the products and $\left<\mathrm{d}' Q\right>/\mathrm{d}t$ is heat rate.
Note that heat is assigned a positive sign for energy absorbed by the system.
To find the heat, the energy input and the stored energy are first determined by their physical interpretations.

Utilizing eq.~(\ref{energy_in_over}), we can find eq.~(\ref{decomp_over}) as in the form of the time derivative calculation of the H-theorem.
It must be noted that a calculation without an additional condition arrives at
\begin{eqnarray}
{\cal I}_J^* 
&\equiv&
k_\mathrm{B}T 
\sum_{\Lambda_{f}} J_x^{\Lambda_{f}}(L_N) \log{\frac{P^{\Lambda_{f}}(L_N)}{[P(L_N)/M^N]}}
\label{infoflow_overd_P},
\end{eqnarray}
instead of eq.~(\ref{infoflow_overd}).
The discrepancy can be compensated for with a condition near the boundary.
Let us consider the difference between eqs.~(\ref{infoflow_overd}) and (\ref{infoflow_overd_P}).
\begin{eqnarray}
\frac{{\cal I}_J -{\cal I}_J^*}{k_\mathrm{B}T}
&=&\sum_{\Lambda_{f}}  J_x^{\Lambda_{f}} \log{\left( \frac{J_x^{\Lambda_{f}}/P^{\Lambda_{f}}}{J_x/P} \right) }.
\label{compare_difinition}
\end{eqnarray}
Here ${\cal I}_J -{\cal I}_J^* \simeq 0$ is expected for the following reasons:
The transition rates around the terminal end are similar, i.e., $W^{\Lambda_{N,\alpha} \rightarrow \Lambda_{f}}(x) \simeq W^{\Lambda'_{N,\beta} \rightarrow \Lambda'_{f}}(x)$ and $W^{\Lambda_{f} \rightarrow \Lambda_{N,\alpha}}(x)\simeq W^{\Lambda_{f}' \rightarrow \Lambda'_{N,\beta}}(x)$, since the state transition around the last region is irrelevant to the bulk concentration and the difference in the detaching process between sequences is appropriately small.
Then, from eq.~(\ref{over_FP}), we find that the effective velocity exhibits only a small difference, i.e., $J_x^{\Lambda_{f}}(L_N)/P^{\Lambda_{f}}(L_N) \simeq J_x^{\Lambda_{f}'}(L_N)/P^{\Lambda_{f}'}(L_N)$, for $\Lambda_{f} \neq \Lambda_{f}'$.
In addition, if the transition rates are the same, approximation ($\simeq$) is replaced by equality ($=$).
Then, ${\cal I}_J^* \simeq {\cal I}_J $ (or ${\cal I}_J^* = {\cal I}_J $) is obtained by substituting $J_x(L_N)/P(L_N)$ for $J_x^{\Lambda_{f}}(L_N)/P^{\Lambda_{f}}(L_N)$ in the right-hand side of eq.~(\ref{compare_difinition}).

\end{document}